\documentclass[useAMS]{mn2e} 
 
\usepackage{graphicx} 
\usepackage{txfonts} 
\newcommand\aj{AJ} 
\newcommand\apj{ApJ} 
\newcommand\apjs{ApJS}       
 
\newcommand\aap{A\&A} 
\newcommand\mnras{MNRAS} 
\newcommand\apjl{ApJ}

\title[]{The {\it HST} Large Programme on NGC\,6752. II. Multiple populations at the bottom of the main sequence probed in NIR
\thanks{Based on observations with the NASA/ESA {\it Hubble Space Telescope}, obtained at the Space Telescope Science Institute, which is operated by AURA, Inc., under NASA contract NAS 5-26555.}  }
\author[A.\,P.\, Milone et al.] 
       {A.\,P.\,Milone$^{1}$,
         A.\,F.\,Marino$^{1}$,
         L.\,R.\,Bedin$^{2}$,
         J.\,Anderson$^{3}$,
         D.\,Apai$^{4,5}$,
         A.\,Bellini$^{3}$,\newauthor
         A.\,Dieball$^{6}$,
         M.\,Salaris$^{7}$,
         M.\,Libralato$^{3}$,
         D.\,Nardiello$^{1,2}$,
         P.\,Bergeron$^{8}$,
         A.\,J.\,Burgasser$^{9}$,\newauthor
         J.\,M.\,Rees$^{4,9}$,
         R.\,M.\,Rich$^{10}$,
         H.\,B.\,Richer$^{11}$
\\ 
$^{1}$ Dipartimento di Fisica e Astronomia ``Galileo Galilei'', Univ. di Padova, Vicolo dell'Osservatorio 3, Padova, IT-35122\\
$^{2}$ Istituto Nazionale di Astrofisica - Osservatorio Astronomico di Padova, Vicolo dell'Osservatorio 5, Padova, IT-35122\\
$^{3}$ Space Telescope Science Institute, 3800 San Martin Drive, Baltimore, MD 21218, USA\\
$^{4}$ Department of Astronomy and Steward Observatory, The University of Arizona, 933 N. Cherry Avenue, Tucson, AZ 85721, USA\\
$^{5}$ Lunar and Planetary Laboratory, The University of Arizona, 1640 E. University Blvd., Tucson, AZ 85721, USA\\
$^{6}$ Helmholtz Institut f\"ur Strahlen-und Kernphysik, University of Bonn, Germany\\
$^{6}$ Astrophysics Research Institute, Liverpool John Moores University,146 Brownlow Hill, Liverpool L3 5RF, UK\\
$^{8}$ D\'epartement de Physique, Universit\'e de Montr\'eal, C.P. 6128, Succ. Centre-Ville, Montr\'eal, QC H3C 3J7, Canada\\
$^{9}$Center for Astrophysics and Space Science, University of California San Diego, La Jolla, CA 92093, USA\\
$^{10}$Department of Physics and Astronomy, UCLA, 430 Portola Plaza, Box 951547, Los Angeles, CA 90095-1547, USA\\
$^{11}$Department of Physics and Astronomy, University of British Columbia, Vancouver, BC, V6T 1Z1, Canada\\
       } 
\begin{document} 
\date{Accepted 2019 January 21. Received 2019 January 21; in original form 2018 December 13} 
\pagerange{\pageref{firstpage}--\pageref{lastpage}} \pubyear{2017} 
 
\maketitle 
\label{firstpage} 
 
\begin{abstract}
  Historically, multiple populations in Globular Clusters (GCs) have been mostly studied from ultraviolet and optical filters down to stars that are more massive than $\sim$0.6$\mathcal{M}_{\odot}$.
  Here we exploit deep near-infrared (NIR) photometry from the {\it Hubble Space Telescope} to investigate multiple populations among M-dwarfs in the GC NGC\,6752.   We discovered that the three main populations (A, B and C), previously observed in the brightest part of the color-magnitude diagram, define three distinct sequences that run from the main-sequence (MS) knee towards the bottom of the MS ($\sim$0.15 ${\mathcal M}_{\odot}$).
 These results, together with similar findings on NGC\,2808, M\,4, and $\omega$\,Centauri, demonstrate that multiple sequences of M-dwarfs are common features of the color-magnitude diagrams of GCs.
The three sequences of low-mass stars in NGC\,6752 are consistent with stellar populations with different oxygen abundances. The range of [O/Fe] needed to reproduce the NIR CMD of NGC\,6752 is similar to the oxygen spread inferred from high-resolution spectroscopy of red-giant branch (RGB) stars.
  The relative numbers of stars in the three populations of M-dwarfs are similar to those derived among RGB and MS stars more massive than $\sim$0.6$\mathcal{M}_{\odot}$.    
As a consequence, the evidence that the properties of multiple populations do not depend on stellar mass is a  constraint for the formation scenarios.  
  
\end{abstract}
\begin{keywords} 
\end{keywords} 
 
\section{Introduction}\label{sec:intro} 
Near-infrared Wide Field Camera 3 (WFC3/NIR) observations have proved very effective in separating and characterizing multiple populations of faint M-dwarfs in NGC\,2808, NGC\,6121 (M\,4) and $\omega$\,Centauri (Milone et al.\,2012a, 2014, 2017a). 

This work is based on deep {\it HST} images of the nearby globular cluster (GC) NGC\,6752 collected as part of the {\it Hubble-Space-Telescope} ({\it HST}\,) program GO-15096  (PI.\,L.\,R.\,Bedin). The main target of the project is the white-dwarf cooling sequence of NGC\,6752, 
which is the subject of separate work by Bedin et al.\,(in preparation, see also Bedin et al.\,2008, 2009, 2010, 2015). In this paper, we exploit parallel  WFC3/NIR observations only to investigate, for the first time, the multiple populations of NGC\,6752 at the bottom of the main sequence (MS). 

NGC\,6752 is one of the most-studied GCs in the context of multiple stellar populations. Optical and Ultraviolet observations, mostly from {\it HST}, revealed that its color-magnitude diagram (CMD) hosts three distinct red-giant branches (RGBs) and MSs, which comprise a population, A, of stars with primordial helium abundance (Y$\sim$0.246) and two stellar populations, B and C, enhanced in helium by $\Delta$Y$\sim$0.01 and 0.03, respectively  (Milone et al.\,2010, 2013, 2018; Dotter et al.\,2015). 

Spectroscopy shows that population-C stars are enhanced in N, Al, Si, and Na and depleted in C, O, and Mg, with respect to the population A, while population-B stars have intermediate chemical composition (e.g.\,Grundahl et al.\,2002; Yong et al.\,2003, 2005, 2013, 2015; Carretta et al.\,2009, 2012, see Table~2 from Milone et al.\,2013 for the average chemical composition of the stars in the three stellar populations of NGC\,6752).
 Multi-band {\it HST} photometry suggests possible star-to-star helium variations among population-A stars (Milone et al.\,2017b, 2018).

 While multiple populations of NGC\,6752 are widely studied along the RGB and the upper MS, the M-dwarf regime is almost unexplored. The only available information on stellar populations among low-mass stars is provided by Dotter et al.\,(2015), who show that the F110W$-$F160W color distribution of NGC\,6752 stars, estimated $\sim$1.5 F160W magnitudes below the MS knee, is much larger than that expected from a simple population.

 The exquisite photometry from the {\it HST} large program on NGC\,6752 allowed us to identify and characterize, for the first time, the three distinct stellar populations of NGC\,6752 at the bottom of the MS.
 The paper is organized as follows. In Section~\ref{sec:data} we describe the dataset, the data reduction and analysis. The NIR CMD of NGC\,6752 is presented in Section~\ref{sec:cmd}, where we also derive the fraction of stars in the three sequences of M-dwarfs and constrain their chemical composition. Finally, Section~\ref{sec:discussion} provides the summary and the discussion of the results. 

\section{Data and data analysis} \label{sec:data}
We used WFC3/NIR on board of {\it HST} to investigate stellar populations in a field located $\sim$4.8 arcmin north-east from the center of NGC\,6752.  
The images are collected between September 7$^{\rm th}$ and September 18$^{\rm th}$, 2018 as part of GO-15096 (PI.\,L.\,R.\,Bedin). All the observations are obtained during 35 orbits of {\it HST}, including 25 and 10 orbits for images in F110W and F160W, respectively. During each orbit we collected one short exposure of 143\,s in multiaccum mode (with instrument parameters NSAMP = 15, SAMP-SEQ = SPARS10)\footnote{Instrument Handbook, Section 7.7\\ http://www.stsci.edu/hst/wfc3/documents/handbooks/currentIHB/c07 ir08.htm} and two long exposures of 1303\,s each (with instrument parameters NSAMP = 14, SAMP-SEQ = SPARS100). All the images are properly dithered.

To derive photometry and astrometry we adapted to WFC3/NIR images the software presented by Anderson et al.\,(2008) for images collected with the Wide Field Channel of the Advanced Camera for Survey (WFC/ACS). In a nutshell, the software performs Point-Spread-Function fitting of all the sources in the field of view. It uses two different approaches that provide optimal results for bright and faint stars, respectively. The fluxes and positions of bright stars are measured in each image separately, and then averaged. Faint stars in a given patch of the sky are fitted simultaneously by using all the images, once transformed in a common reference frame. To do this, the software exploits library PSFs and the geometric-distortion solution by Jay Anderson\footnote{http://www.stsci.edu/\~\,jayander/WFC3/}. 

We calibrated our photometry to the Vega-magnitude system by following the procedure described by Bedin et al.\,(2005) and adopting the photometric zero points provided by STScI web page\footnote{http://www.stsci.edu/wfc3/phot\_zp\_lbn} for WFC3/NIR.
We used various diagnostics of the photometric and astrometric qualities to select a sample of relatively isolated stars that are well fitted by the PSF. These diagnostics comprise the position and magnitude rms, the fraction of flux in the aperture due to neighbours and the quality of the PSF fit.
 We plotted each parameter as a function of the stellar magnitude and verified that most stars follow a clear trend. Outliers include stars with poor astrometry and photometry and are excluded from our investigation of multiple populations. We refer to papers by Milone et al.\,(2009) and Bedin et al.\,(2009) for details. 

 To account for the spatial variations of the photometric zero point across the field that are due to small inaccuracies in the PSF model and in the background determination we used the procedure by Milone et al.\,(2012b, see their Section.~3.2). Briefly, we derived the MS fiducial line and examined the color residuals relative to this sequence.  We corrected the color of each star by the median color residual of its 45 best-measured neighbors. These corrections are typically smaller than 0.007 mag, and never exceed 0.015 mag.

\subsection{Artificial stars}\label{subsec:ASs}
Artificial stars (ASs) are used to estimate photometric errors and to derive the completeness level of our sample and are run by using the procedure by Anderson et al.\,(2008). Briefly, we generated a catalog including the coordinates and the F110W and F160W magnitudes of 100,000 stars randomly distributed along the field of view. The ASs have instrumental magnitudes, $-2.5$log$_{10}$(flux), ranging from $-10.0$ to 0.0 in the F160W band, while the corresponding F110W magnitudes are derived from the fiducial lines of the three sequences. 

The ASs are reduced by following the same procedure and by exploiting the same computer programs by Anderson and collaborators that we used for real stars. We calculated for ASs the same diagnostics of the photometric and astrometric quality derived for real stars and we selected a sample of well-measured ASs by using the same criteria described above for real stars.

To calculate the completeness that corresponds to the position and luminosity of each star we divided the field of view into five concentric annuli centred on the cluster center, and within each annulus we examined the AS results in five magnitude bins.
We calculated the ratio of the well-measured recovered ASs to the input ASs in each of these 5$\times$5 grid points.
 The completeness associated to each star in the field of view is estimated by linearly interpolating among these grid points. We performed distinct completeness calculation for stars in each population.

\section{The NIR CMD of NGC\,6752}\label{sec:cmd}
\begin{centering} 
\begin{figure} 
 \includegraphics[width=9.0cm]{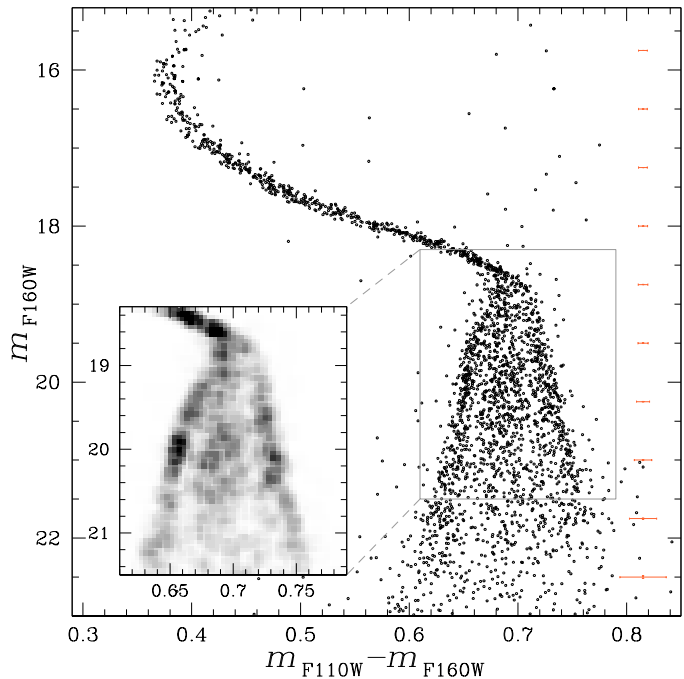}
 \caption{NIR CMD of NGC\,6752. The Hess diagram highlights the region of the CMD where the three sequences are more evident.} 
 \label{fig:cmd} 
\end{figure} 
\end{centering}

The $m_{\rm F160W}$ vs.\, $m_{\rm F110W}-m_{\rm F160W}$ CMD of all the stars in the WFC3/NIR field of view is plotted in Fig.~\ref{fig:cmd}.
A visual inspection of this figure reveals that the MS is narrow and well defined in the magnitude interval between the MS turn off and the MS knee, where the $m_{\rm F110W}-m_{\rm F160W}$ color width is nearly consistent with the broadening expected from observational errors. 
In contrast, the color distribution of MS stars fainter than the MS knee is significantly wider than what we expect from photometric errors alone, and the MS width ranges from $\sim 0.04$ mag for $m_{\rm F160W} \sim 19$ to more than 0.15 mag  around $m_{\rm F160W} = 23$. The Hess diagram plotted in the inset reveals three MSs of low mass stars that run from the MS knee towards the bottom of the MS.

\begin{centering} 
\begin{figure*} 
 \includegraphics[width=13.5cm]{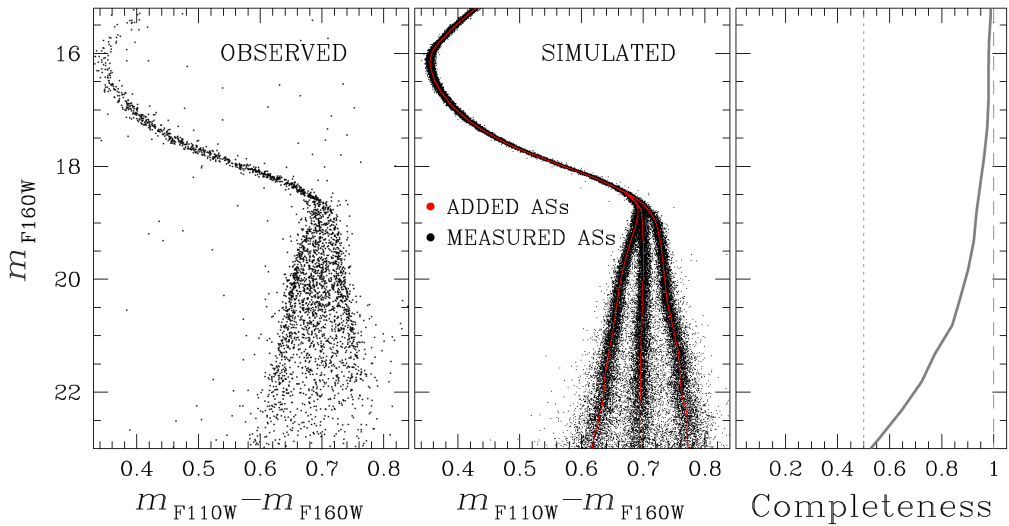}
 \caption{\textit{Left panel.} Reproduction of the observed CMD plotted in Fig.~\ref{fig:cmd}. \textit{Middle panel.} Simulated CMD derived from ASs. Red points are derived from the magnitudes of ASs in the input catalogue, while the measured magnitudes and colors of ASs are colored black. \textit{Right panel.} Average completeness as a function of the F160W magnitude.} 
 \label{fig:ASs} 
\end{figure*} 
\end{centering}

In Fig.~\ref{fig:ASs} we compare the observed CMD of NGC\,6752 with the simulated CMD derived from ASs. As discussed in Sect.~\ref{subsec:ASs}, the latter comprises three stellar populations that are distributed along the fiducials of the three MSs and are colored red in Fig.~\ref{fig:ASs}. The comparison between the observations and the measured ASs, which are represented with black points, corroborates the conclusion that NGC\,6752 host three main populations of M dwarfs. Finally, we plot in the right panel of Fig.~\ref{fig:ASs} the average stellar completeness against $m_{\rm F160W}$. 

\subsection{Comparison with theory}\label{subsec:theory}

\begin{centering} 
\begin{figure*} 
 \includegraphics[height=7.5cm]{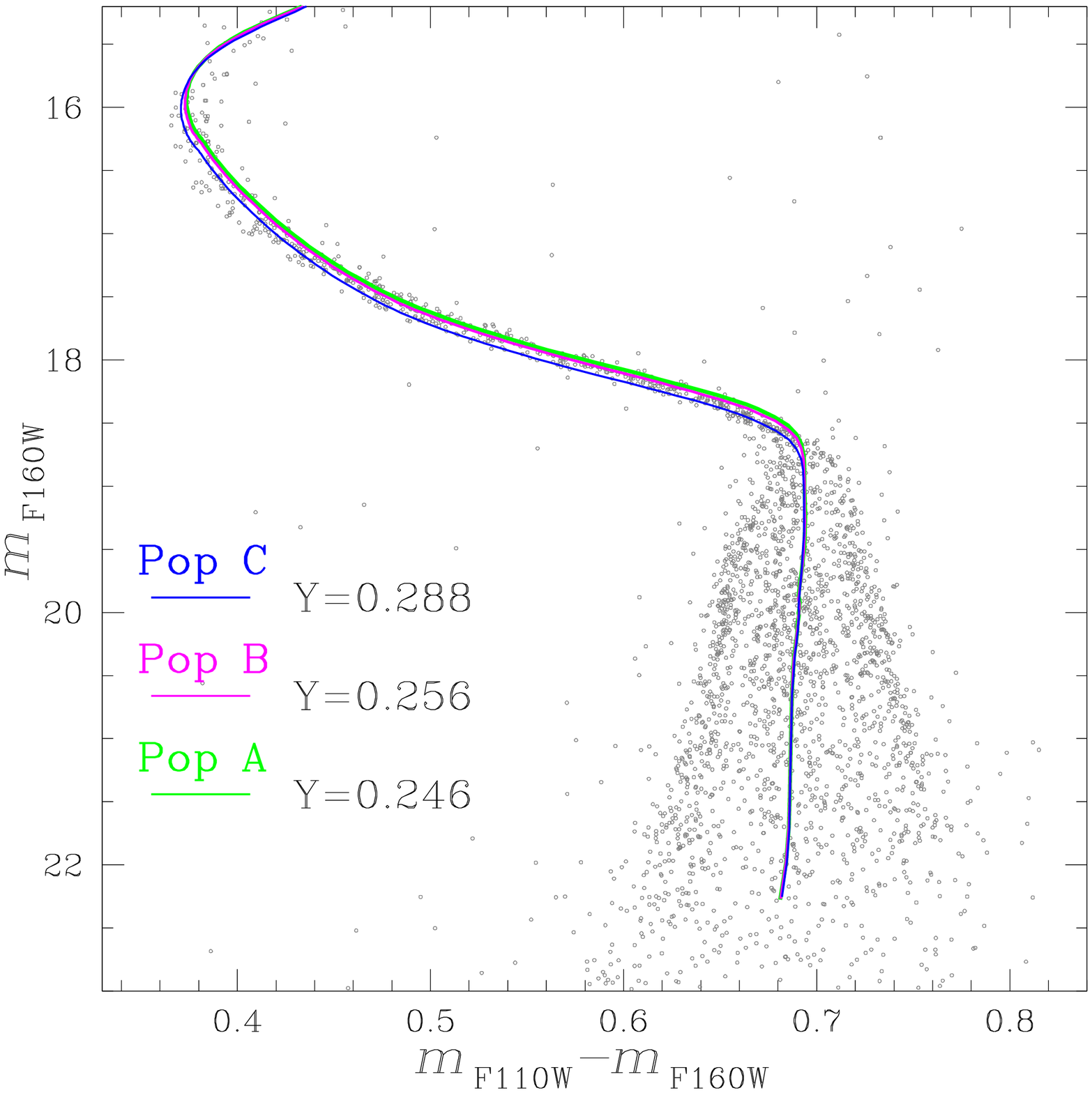}
 \includegraphics[height=7.5cm]{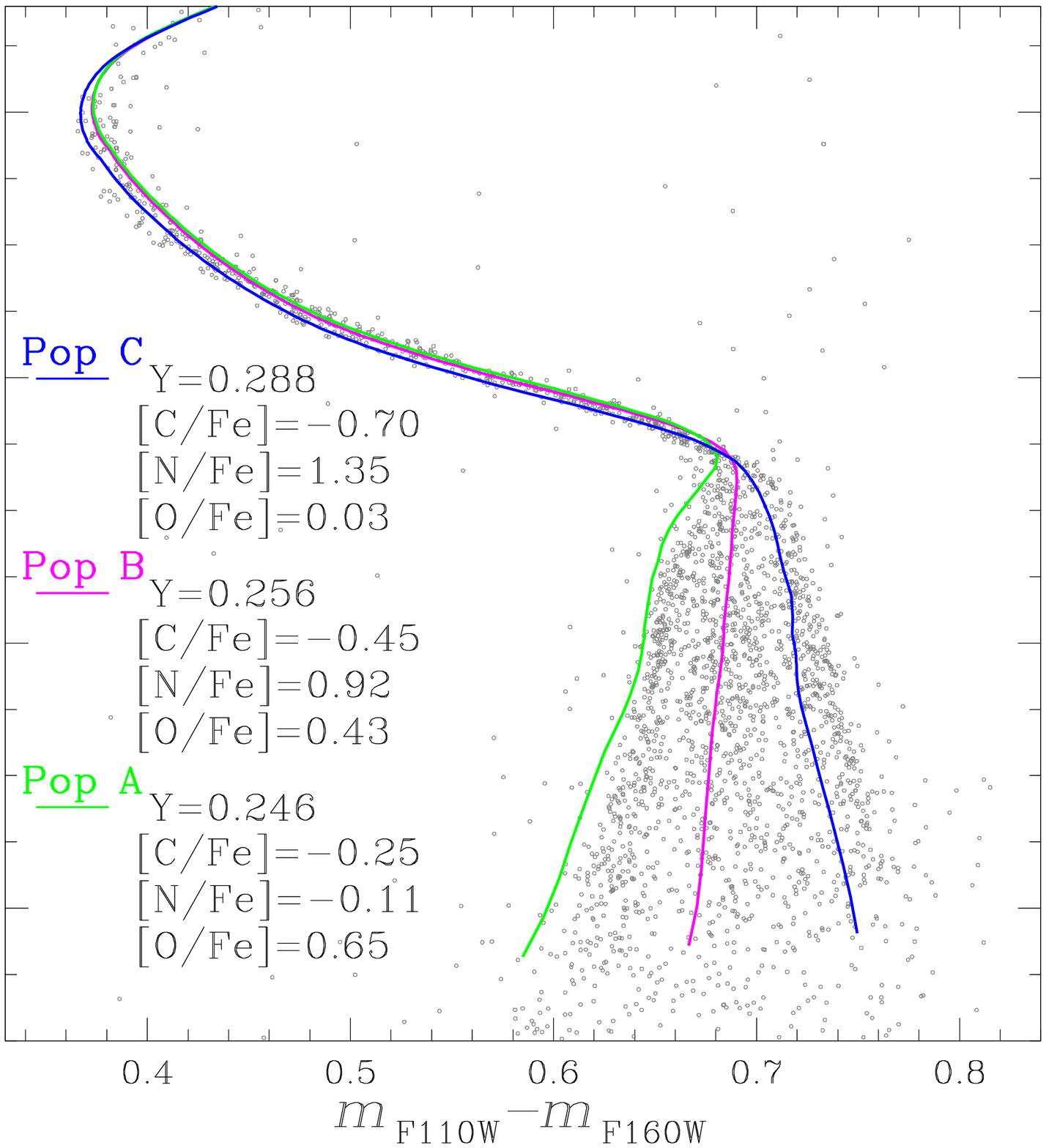}
 \caption{Dartmouth isochrones with different helium abundance overimosed on the  CMD of Fig.~\ref{fig:cmd}. In the left-panel,  where we assumed the same abundances of C, N and O, the three isochrones have similar F110W$-$F160W colors and are difficult to be distinguished for $m_{\rm F160W} \gtrsim 18.5$. Right-panel isochrones account for the C, N and O abundances of population A, B and C as inferred from spectroscopy and quoted in the figure. The comparison between the right-panel isochrones and the observations demonstrate that the observed blue, middle and red MS correspond to population A, B, and C, respectively. See text for details.} 
 \label{fig:iso} 
\end{figure*} 
\end{centering}

To characterize the three MSs of NGC\,6752 we compared the observed CMD with isochrones from the Dartmouth database (Dotter et al.\,2008) as shown in Fig.~\ref{fig:iso}. We assumed [Fe/H]=$-1.61$  from Yong et al.\,(2005), [$\alpha$/Fe]=0.4 and adopted the age, t=12.5 Gyr, distance modulus, (m$-$M)$_{0}$=12.95, and reddening E(B$-$V)=0.07 that provide the best fit with the data. The reddening has been converted into absorption in the F110W and F160W bands by using the relations: A$_{\rm F110W}$=1.016$\cdot$E(B$-$V) and A$_{\rm F160W}$=0.629$\cdot$E(B$-$V) kindly provided by Aaron Dotter (private communication).  
 The values of reddening and distance are in agreement with those listed in the Harris\,(1996, updated as in 2010) catalog, while the age is consistent with previous estimate by Dotter et al.\,(2010). 
 We assumed different helium contents for the three populations of NGC\,6752 of Y=0.246, Y=0.256, and Y=0.288, which are the values inferred by Milone et al.\,(2013, 2018) from multiple MSs and RGBs.
 
 The best-fit isochrones are overimposed on the observed CMD plotted in the left panel of Fig.~\ref{fig:iso} and provide a poor fit with the MS region below the knee. This fact is quite expected because the F110W$-$F160W color of M-dwarfs are significantly affected by the content of oxygen and we assumed the same value of [O/Fe] for the three populations, in contrast with what is inferred from spectroscopy (e.g.\,Grundahl et al.\,2002; Yong et al.\,2003, 2005, 2013; Carretta et al.\,2009).
 
To account for light-element abundance variations, we identified a series of fifteen points along the MS that span the magnitude interval between $m_{\rm F160W}=15.5$ and $m_{\rm F160W}=23.0$  in steps of 0.5 magnitudes and calculated for each of them four synthetic spectra.  To calculate all the spectra we assumed [Fe/H]=$-1.61$, [$\alpha$/Fe]=0.4 and the values of effective temperature and gravity inferred by the best-fit isochrones and a microturbolent velocity of 2 km s$^{-1}$.
 Specifically for each population, we calculated a reference synthetic spectrum with solar carbon and nitrogen abundance and with [O/Fe]=0.4 and a comparison spectra with the corresponding average abundances of C, N and O derived from spectroscopy. Specifically, we used ([C/Fe], [N/Fe], [O/Fe])=($-$0.25, $-$0.11, 0.65), ($-$0.45, 0.92, 0.43) and ($-$0.70, 1.35, 0.03) for population A, B, and C, respectively (see Table~2 from Milone et al.\,2013 and Yong et al.\,2005, 2013, 2015). 

 Synthetic spectra are generated with SYNTHE (Kurucz \& Avrett 1981) over the wavelength range between 8,500 and 19,000 $\AA$ and by using model atmospheres calculated by using the ATLAS12 code, which was developed by Robert Kurucz (e.g.\,Kurucz 1970, 1993, 2005; Castelli 2005) and ported to Linux by Sbordone et al.\,(2004, 2007). We included molecular line lists for CO, C$_{2}$, CN, OH, MgH, SiH, H$_{2}$O, TiO, VO, ZrO from Partridge \& Schwenke\,(1997), Schwenke (1998).
 Each synthetic spectrum has been convolved with the transmission curves of the F110W and F160W WFC3/NIR filters to derive the corresponding fluxes. Finally, we calculated the difference between the F110W and F160W magnitudes derived from each comparison spectrum and the corresponding reference spectrum ($\delta$m). The blue, green and red isochrones plotted in the right panel of Fig.~\ref{fig:iso} have been derived by adding $\delta$m to the corresponding isochrones plotted in the left panel.
 
 The isochrones that account for the abundances of He, C, N, O of the distinct populations qualitatively reproduce the observations and show that the blue, middle, and red MSs correspond to the populations A, B and C defined by Milone et al.\,(2013). As expected, the isochrones of populations A and C are qualitatively similar to those calculated by Dotter et al.\,(2015, see their Figures~9 and 12) who used ATLAS\,12 and similar chemical compositions for populations A and C.
 
 The reason why these isochrones provide a better fit with the observed CMD is that the split MS is a natural consequence of the fact that population-A stars have higher oxygen abundance than population-C stars.
Indeed, the F160W band is heavily affected by absorption from H$_{2}$O molecules, which is stronger in MS-A stars than in MS-C stars, while the F110W filter is almost unaffected by the oxygen abundance. As a consequence, MS-A stars have fainter F160W magnitudes and bluer F110W$-$F160W colors  than MS-C stars with the same luminosity. MS-B stars have intermediate properties.
 
\subsection{The three stellar populations of NGC\,6752}

\begin{centering} 
\begin{figure*} 
 \includegraphics[width=11.0cm]{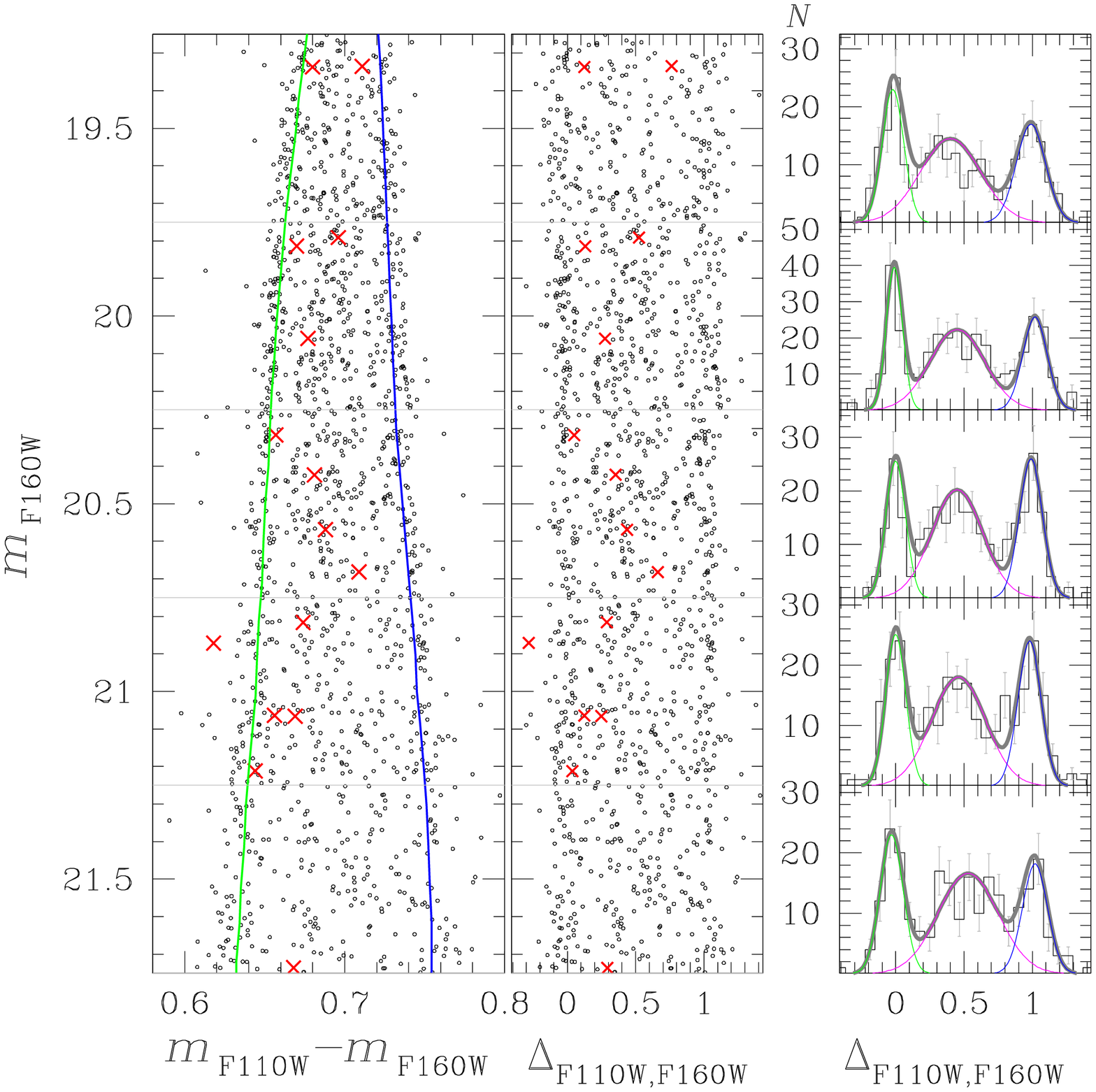}
 \caption{This figure illustrates the procedure that we used to estimate the fraction of MS-A, MS-B, and MS-C stars with respect to the total number of MS stars. Left panel is a zoom of the CMD plotted in Fig.~\ref{fig:cmd} in the region where the three MSs are more-clearly distinguishable. The green and blue lines are the fiducials of MS-A and MS-C respectively, and are used to derive the verticalized $m_{\rm F160W}$ vs.\,$\Delta_{\rm F110W, F160W}$ diagram plotted in the middle panel. The red crosses mark field stars that we simulated by using the population-synthesis code Trilegal (Girardi et al.\,2005). Right panels show the $\Delta_{\rm F110W, F160W}$ histogram distribution of cluster stars in the five luminosity intervals indicated by the gray lines plotted in the left and middle panels. The gray-tick lines overimposed on each histogram are the best-fit three-Gaussian functions, whose components are represented with thin green, magenta and blue lines. Clearly, the fraction of MS-A, MS-B, and MS-C stars are similar in the five magnitude intervals. Moreover, MS-B is more-broadened than MS-A and MS-C thus indicating that its stars are not chemically homogeneous. See text for details. } 
 \label{fig:pratio} 
\end{figure*} 
\end{centering} 

To derive the relative numbers of stars in each MS, we followed the procedure illustrated in Fig.~\ref{fig:pratio}.
The left panel of Fig.~\ref{fig:pratio} is a zoom of Fig.~\ref{fig:cmd} around the region of the NIR CMD where the triple sequence is clearly distinguishable and the blue and red lines overimposed on the CMD are the fiducials of MS-A and MS-C, respectively. To derive the red (blue) fiducial we applied a procedure based on the naive estimator method (Silverman 1986). In a nutshell, we first defined a series of magnitude bins of width $\nu=0.25$ mag over a grid of $N$ points separated by steps of fixed magnitude ($s=\nu/5$). Then, we calculated the medians of the $m_{\rm F110W}-m_{\rm F160W}$ color distribution and of the F160W magnitude distribution of all the MS-A (MS-C) stars in each bin. Finally, we smoothed the median points by using the boxcar average smoothing, where each point has been replaced by the average of the three nearby points.
The fiducial lines are used to derive the verticalized diagram plotted in the middle panel of Fig.~\ref{fig:pratio}, where the abscissa is calculated as:
\begin{equation}
\Delta_{\rm F110W,F160W}= [(X-X_{\rm red~fiducial})/(X_{\rm blue~fiducial}-X_{\rm red~fiducial})]-1 
\end{equation}
with $X=m_{\rm F110W}-m_{\rm F160W}$.
To account for the contamination from foreground and background field stars, we used the population-synthesis code Trilegal (Girardi et al.\,2005) to simulate the $m_{\rm F160W}$ vs.\,$m_{\rm F110W}-m_{\rm F160W}$ CMD of Milky-Way stars  in a square-degree field of view in the direction of NGC\,6752.
The red crosses overimposed to the diagrams plotted in the left- and middle-panel diagrams of Fig.~\ref{fig:pratio} are the stars expected in a region with the same area of the analyzed WFC3/NIR field of view and are randomly extracted from the entire sample of simulated field stars. 
 This figure immediately reveals that the contamination of field stars is negligible and that the triple sequence is not due field stars. 

 The $\Delta_{\rm F110W,F160W}$ histogram distribution of stars in five magnitude intervals is plotted in the right panel of Fig.~\ref{fig:pratio}.
 To derive each histogram we accounted for stellar completeness and subtracted the field stars.  The black continuous lines overimposed on the histograms are the best-fit least-squares three-Gaussian functions, whose components are colored green, magenta and blue.
 From the average of the population ratio obtained in the five magnitude intervals we obtain that the MS-A, MS-B and MS-C host 26.3$\pm$1.4\%, 46.9$\pm$1.3\% and 26.8$\pm$2.0\%, respectively, of the total number of MS stars. The uncertainties are estimated as the ratio between the r.m.s. of the five population-ratio measurements divided by the square root of four.
 The fractions of MS-A, MS-B and MS-C stars with respect to the total number of MS stars are estimated from the areas of the corresponding Gaussians. Results are listed in Table~1, where we also provide the dispersion of each Gaussian component. Noticeably, the MS-B is significantly wider than the other two sequences, thus indicating that its stars are not chemically homogeneous.

\begin{table*}
 \caption{Fraction of MS-A ($f_{\rm MS-A}$), MS-B ($f_{\rm MS-A}$) and MS-C ($f_{\rm MS-A}$) stars with respect to the total number of MS stars inferred in five $m_{\rm F160W}$ magnitude bis from the best-fit Gaussian functions of Fig.~\ref{fig:pratio}. We also provide the dispersions of the best-fit Gaussians ($\sigma_{\rm MS-A, MS-B, MS-C}$) and the total number of stars ($N$) in each magnitude bin.}
 \begin{tabular}{cccccccc}
   \hline
   \hline
$m_{\rm F160W}$ bin &  $N$     &  $f_{\rm MS-A}$   & $\sigma_{\rm MS-A}$ & $f_{\rm MS-B}$   & $\sigma_{\rm MS-B}$ & $f_{\rm MS-C}$   & $\sigma_{\rm MS-C}$ \\ 
   \hline
    21.25-21.75 &   248    & 0.276$\pm$0.028     &     0.060         & 0.486$\pm$0.032   &     0.155         &   0.238$\pm$0.027   &    0.065          \\
    20.75-21.25 &   280    & 0.255$\pm$0.026     &     0.052         & 0.464$\pm$0.030   &     0.140         &   0.281$\pm$0.027   &    0.061          \\
    20.25-20.75 &   296    & 0.243$\pm$0.026     &     0.051         & 0.467$\pm$0.029   &     0.132         &   0.290$\pm$0.027   &    0.060          \\   
    19.75-20.25 &   341    & 0.265$\pm$0.024     &     0.042         & 0.476$\pm$0.028   &     0.142         &   0.259$\pm$0.024   &    0.063          \\
    19.25-19.75 &   256    & 0.274$\pm$0.029     &     0.056         & 0.452$\pm$0.031   &     0.156         &   0.274$\pm$0.029   &    0.076          \\
   \hline
 \end{tabular}\\
 \label{tab:data}
 \end{table*}
  
\subsection{The radial distribution of multiple populations}
  In past years, the multiple populations of NGC\,6752 have been investigated at different radial distances from the cluster center. 
  Milone et al.\,(2013, 2017) identified the three main populations of NGC\,6752 within six arcmin from the cluster center and found that populations A, B and C comprise about 25\%, 45\% and 30\% of the total number of cluster stars.
  Nardiello et al.\,(2015) extended the investigation at larger radial distances and concluded that the MS-A hosts 26$\pm$4\% of the total number of MS stars in the region between $\sim$5.9 and 17.9 arcmin from the cluster center. Both papers by Milone et al.\,(2013) and Nardiello et al.\,(2015) conclude that there is no evidence for any radial gradient of the different stellar populations. 
Similarly, Lee\,(2018) analyzed multiple populations in NGC\,6752 by using Ca-CN photometry and confirmed that  the two main groups of CN-weak and CN-strong stars that he identified along the RGB follow the same radial distribution within $\sim$8.2 arcmin from the cluster center.
  
  In the upper panel of Fig.~\ref{fig:pratio1} we plot the ratio of population-A, B, and C stars against the radial distance from the cluster center by using results from literature and from this paper. The present investigation, which is based on a field with radial distance between 3.3 and 6.6 arcmin from the center of NGC\,6752, corroborates the conclusion that the three populations of NGC\,6752 are consistent with a flat radial distribution. 
  
\subsection{Dependence of multiple populations from stellar mass}
Published studies on multiple populations in NGC\,6752 are mostly focused on RGB and bright MS stars, which are more massive than $\sim$0.6$\mathcal{M}_{\odot}$. 
 Specifically, results by Milone et al.\,(2013, 2017) are based either on {\it HST} photometry of MS stars with $19.05<m_{\rm F814W}<20.15$ or on photometry of RGB stars from {\it HST} and from the Str${\rm \ddot{o}}$mgren catalogue by Grundahl et al.\,(2002) obtained from New-Technology-Telescope data. The conclusion by Nardiello et al.\,(2015) are inferred from photometry of MS stars with $19.25<V<20.51$.

\begin{centering} 
\begin{figure} 
 \includegraphics[width=8.7cm]{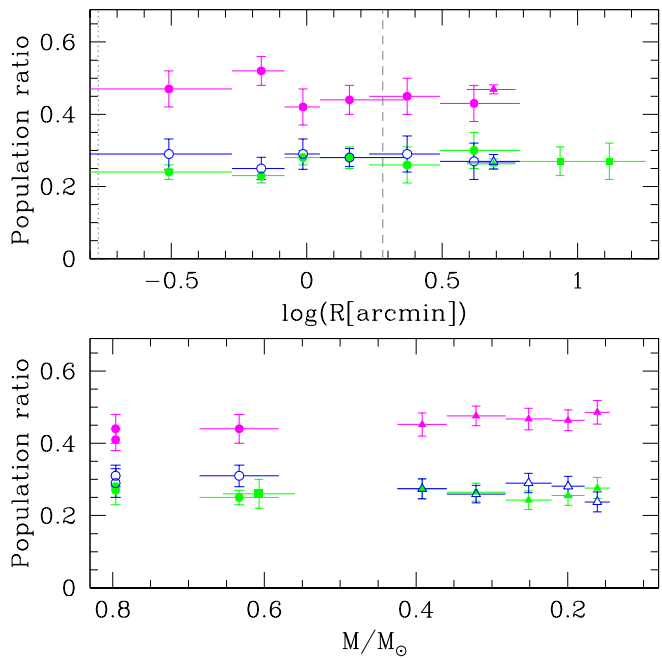}
 \caption{Fractions of stars in the populations A (green), B (magenta), and C (blue) against radial distance from the cluster center (upper panel) and stellar mass (lower panel). The fractions of stars more massive than 0.5 solar masses are taken from Milone et al.\,(2013, 2017  circles) and Nardiello et al.\,(2015, squares), while the results for low-mass stars are derived in this paper and are represented with triangles. This figure suggests that there is no evidence for strong dependence between the population ratios and neither stellar mass nor radial distance.} 
 \label{fig:pratio1} 
\end{figure} 
\end{centering}

The lower panel of Fig.~\ref{fig:pratio1} shows the fraction of stars in the distinct populations as a function of the stellar mass, where the stellar masses are derived by using the mass-luminosity relation provided by the best-fit isochrones derived in Sect.~\ref{subsec:theory}. Results are consistent with constant fraction of population-A, -B and -C stars over the analyzed interval of $\sim0.15$-$0.80$ solar masses.
 
\section{Summary and discussion}\label{sec:discussion}
 In the context of the {\it HST} Large program on NGC\,6752 (GO-15096), we analyzed deep WFC3/NIR photometry to investigate multiple stellar populations among M-dwarfs in NGC\,6752. 
 The $m_{\rm F160W}$ vs.\,$m_{\rm F110W}-m_{\rm F160W}$ CMD of NGC\,6752 reveals that the MS splits into three distinct sequences below the MS knee.

  The comparison between the observations and isochrones that account for the chemical composition of stars in NGC\,6752, demonstrates that the three sequences correspond to the three populations (A, B and C) identified along the brightest part of the MS, the SGB, and the RGB by Milone et al.\,(2013, see also Yong et al.\,2005, 2013).

 The MS-B is significantly wider than both MS-A and MS-C. 
 The fact that its stars exhibit a color spread that is larger than the broadening expected from photometric uncertainties alone, demonstrates that MS-B is not chemically homogeneous.
  
 We find that the MS-A, MS-B and MS-C host 26.3$\pm$1.4\%, 46.9$\pm$1.3\% and 26.8$\pm$2.0\%, respectively, of the total number of MS stars. These results, that we derived in a field of view located about 4.8 arcmin north-east from the cluster center, are similar to the relative numbers of stars in the population-A, B, and C estimated at various radial distances (Milone et al.\,2013; Nardiello et al.\,2015; Lee 2018). Hence, we confirm that there is no evidence for any strong radial gradients within $\sim$16 arcmin (i.\,e.\,$\sim$8.4 half-light radii) from the center of NGC\,6752.

 The relative numbers of population-A, population-B, and population-C stars are almost constant along the entire interval that ranges from $\sim$0.15 to 0.80 solar masses. Moreover, the range of [O/Fe] needed to reproduce the color broadening of stars at the bottom of the MS is similar to that inferred from spectroscopy of more-massive RGB stars.
Noticeably, Milone et al.\,(2014) obtained similar conclusion for M\,4, where the relative numbers of stars in the two stellar populations of this cluster and the oxygen abundances inferred from M-dwarfs are similar to the corresponding quantities measured among RGB and more-massive MS stars (Marino et al.\,2008, 2011, 2017; Nardiello et al.\,2015; Milone et al.\,2017). 
These facts suggest that the properties of the multiple populations of GCs do not depend significantly on the stellar mass in the $\sim$0.15-0.80 $\mathcal{M}_{\odot}$ range, and provide new constraints to the scenarios for the formation of multiple populations.

According to some scenarios, GCs have experienced multiple episodes of star formation, where the material ejected by relatively massive first-generation (1G) stars produces a cooling flow, rapidly collects in the innermost regions of the cluster and forms centrally-concentrated second stellar generations (2G, e.g.\,Ventura et al.\,2001; Decressin et al.\,2007; De Mink et al.\,2009; D'Ercole et al.\,2010; Denissenkov et al.\,2014; D'Antona et al.\,2016).
It has been suggested that the proto GCs should have been substantially (by a factor of about ten) more massive at birth.
 This conclusion comes from the evidence that 1G stars are the minority population in most GCs and that only a fraction of 1G stars is turned into 2G stars.
 As a consequence, the proto GCs should have lost the majority of their 1G stars into the Galactic halo thus providing a major contribution to the assembly of the Galaxy. In this case, 1G and 2G stars could form in environments with different densities and follow a different dynamical evolution during the early phases of cluster formation.

 The present-day 1G/2G ratio among stars with different masses, depends on the initial mass functions of the distinct stellar populations and on their dynamical evolution (see Vesperini et al.\,2018). As a consequence, the evidence that the population ratios are almost constant along the $\sim$0.15-0.80 $\mathcal{M}_{\rm \odot}$ mass interval provides strong constraints to the early phases of GC formation. 
 
 Alternative scenarios of formation of multiple populations in GCs suggest that there would have been only one episode of star formation, but a fraction of stars would have successively accreted material processed and polluted by massive stars of the same generation (e.g.\,Bastian et al.\,2013; Gieles et al.\,2018). In these scenarios, the amount of accreted material would depend on the stellar mass.
 
 Our work shows that the oxygen variation needed to reproduce the observed colors of multiple MSs of low-mass stars is consistent with the range of [O/Fe] inferred from spectroscopy of more-massive RGB stars. As a consequence, in any scenario based on accretion onto already-existing stars, the rate of accreated material should be proportional to the stellar mass.
 As an example, our results exclude a Bondi accretion, where the amount of accreted material is proportional to the square of the stellar mass and low-mass stars would be inefficient to accreate polluted material.

 In general, the discovery of multiple MSs of faint MS stars in NGC\,6752, together with similar findings on NGC\,2808, M\,4 and $\omega$\,Cen, indicate that multiple sequences of M-dwarfs are common features of GCs.
  The evidence that stellar populations with different chemical compositions are also present in M-dwarfs, which are fully-convective stars, demonstrates that the chemical composition of multiple populations is not a product of stellar evolution.

\section*{acknowledgments} 
\small 
We thank the referee for her/his work that improved the manuscript. This paper has received funding from the European Research Council (ERC) under the European Union's Horizon 2020 research innovation programme (Grant Agreement ERC-StG 2016, No 716082 `GALFOR', PI: Milone), and the European Union's Horizon 2020 research and innovation programme under the Marie Sklodowska-Curie (Grant Agreement No 797100, beneficiary: Marino). APM acknowledges support from MIUR through the the FARE project R164RM93XW ‘SEMPLICE’. 
JA, AB and ML acknowledge support from STScI grant GO-15096.
\bibliographystyle{aa}

\end{document}